\documentstyle[12pt,aasms]{article}
\newcommand{\skp}{\mbox{  }\\}

\newcommand{\etal}{{\it et al}.\ }

\newcommand{\gta}{\stackrel{>}{\sim}}

\begin{document}

\title{The Long Period AM Her-like Cataclysmic Variable RXJ051541+0104.6}
\author{Frederick M. Walter, Scott J. Wolk\altaffilmark{1},
        and Nancy R. Adams}
\affil{Earth and Space Sciences Department\\
 State University of New York\\
 Stony Brook NY 11794-2100\\
 I: fwalter@astro.sunysb.edu}
\altaffiltext{1}{Visiting Astronomer, Kitt Peak National
     Observatory, National Optical Astronomy Observatories, which is
     operated by the Association of Universities for Research in Astronomy,
     Inc., under contract with the National Science Foundation.}

\begin{abstract}
We report the discovery of a new catacysmic variable system, RXJ051541+0104.6.
The optical spectrum has a blue continuum with superposed H~I and He~I and II
emission lines. The soft X-ray spectrum is well fit with a 50~eV black body.
The X-ray and optical data are suggestive of an AM~Herculis system. The X-ray
light curve shows extreme variability on timescales of seconds, and suggests an
orbital period of order 8~hours, nearly twice that of the longest catalogued
AM~Her period. When bright, the X-ray light curve breaks up into a series of
discrete bursts, which may be due to accretion of dense blobs of material of
about 10$^{17}$~gm mass.

\end{abstract}
Subject Headings: stars: cataclysmic variables -- X-rays: stars

\section{Introduction}

As part of a program to study the spatial distribution of low mass pre-main
sequence stars in the Orion OB1 association, we obtained the ROSAT PSPC
observation RP200930. This image is centered at 5$^h$14$^m$24.0$^s$ +1$^o$42'0"
(J2000), to the west of the Ori~OB1a association, and was meant to serve as a
control field, far from the region where we expected to find low mass
association members. The standard SASS processing yielded 37 X-ray sources. In
the course of obtaining optical spectra of the stellar counterparts of these
X-ray sources, we discovered a previously unknown cataclysmic variable with
extreme X-ray variability.

Cataclysmic variables are semi-detached binary systems, with accretion from a
low mass non-degenerate star onto a white dwarf. The cataclysmic variables come
in a number of classes, defined by accretion rates and accretion geometry
(e.g., C\'ordova 1993). Among these are the various types of novae and
nova-like systems and the magnetized systems, the polars and intermediate
polars. Polars and intermediate polars are often discovered by virtue of their
X-ray emission.

The polars, or AM~Her systems (e.g., Cropper 1990), are strongly magnetized
systems with the white dwarf rotating synchronously on the orbital period.
There is no evidence for accretion disks in polars: the accretion is along
magnetic field lines onto the magnetic poles of the white dwarf. Orbital
periods are generally less than 2 hours, but have periods up to 4.6~hours
(RXJ1313-32; Ritter \& Kolb 1993). Polars are strong soft X-ray sources, and 20
such systems have been discovered in the ROSAT all-sky survey (e.g., Beuermann
\& Schwope 1994), more than doubling the previously-known population. The
intermediate polars, or DQ~Her systems (e.g., Patterson 1994), are also
magnetized, but the white dwarf periods are much shorter than the orbital
periods, presumably because the magnetic fields are too weak to enforce
synchronization. Accretion is through a disk. Periods tend to be longer than
for the polars, and the X-ray emission tends to be dominated by a hard
bremsstrahlung component.

This new cataclysmic variable exhibits several puzzling characteristics, which
we discuss below. The X-ray light curve resembles that of a single-poled
polar\footnote{We use this term to describe a polar with an X-ray light curve
similar to that of VV~Pup, where there are two magnetic poles but only the
principle pole is occulted by the body of the white dwarf.}, yet the period is
much longer than any known polar. The soft X-ray and optical spectra are more
similar to those of polars than of intermediate polars. The X-ray emission is
highly variable; when bright, the emission breaks up into a discontinuous
series of discrete bursts. This behavior is unlike that seen in any cataclysmic
variable discussed to date, but is expected if the accretion consists not of a
smooth flow but of discrete blobs of material (e.g., Frank, King, \& Lasota
1988). This object may provide clues to the evolution of cataclysmic variables,
and the relation of the polars to the intermediate polars.

\section{X-ray Observations}

The X-ray data were obtained using the {\it R\"ontgen Satellite} (ROSAT;
Tr\"umper 1983; Pfeffermann \etal 1987; Aschenbach 1988). Observation RP200930
was obtained in two segments; 3.7~Ksec between 1992 September 11 and~19, and
6.1~Ksec on 1993 February 27 (observation times are provided in
Table~\ref{tbl-x}). The total exposure time was 9779~sec. The object,
RXJ051541+0104.6, was detected in both segments. It lies 42~arcmin off-axis.
Because of the large off-axis distance, the SASS ML source detection algorithm
was not run, and no formal uncertainties on the source location were provided.
The mean X-ray source position, weighted by the number of counts in the source,
is 5$^h$15$^m$40.3$^s$ +1$^o$4'33.9" (J2000). No catalogued X-ray source or
close binary system is known at this location.

We analyzed these data using the RX package, an IDL-based ROSAT data analysis
system (Walter 1994). We extracted the source counts within a circle of
225~arcsec radius centered on the source. The background was extracted from
within an annulus between 225 and 675~arcsec from the source, excluding a wedge
containing another X-ray source. The extraction areas were normalized using the
PSPC exposure map. The X-ray source is fairly bright, with a mean
vignetting-corrected count rate of 0.21$\pm$0.006 counts~s$^{-1}$ in the PSPC.
A net total of 1028 source counts were extracted.

The source also appears at the edge of a second image, RP700422, 56~arcmin
off-axis. This is a day-long pointing obtained while ROSAT was in the
reduced-pointing mode. The image contains 24.5~Ksec of data in an 85.7~Ksec
interval. We extracted 2972 photons from this source using a 3600~arcsec circle
and an 18~arcmin radius background annulus. The source is visibly truncated by
the edge of the detector, and wobbles out of the field of view about 100 of
every 400 seconds, so we cannot place much confidence in either short-term
variability (less than the typical 1000~sec integration each orbit) or the
spectrum. The background light curve shows strong soft enhancements toward the
ends of many intervals. We have not edited these data out because they have no
significant effect on the background-subtracted light curve, and because we did
not attempt a spectral analysis.

\subsection{X-ray Variability}

The light curve from observation RP700422 is shown in Figure~1. The data show a
repeating pattern suggestive of a period of about 8~hours. Observations from
low Earth orbit can suffer from aliasing with the $\sim$94~minute spacecraft
orbit. During RP700422, observations were obtained on nearly every orbit, for
about 3000 seconds (often in two segments, with a central gap due to SAA
passages). Observation RP200930 contains two 3000 second observation intervals
(OBIs). The source is seen to be low (count rate less than the mean rate) about
60\% of the time, and high the remaining time. If the light curve is
repeatable, then we can rule out any periods less than about 2.1 hours because
we see no significant differences in mean activity level within any single
orbit. The best fit period, from folding the RP700422 light curve on itself, is
8.05$\pm$0.1~hours (Figure~2). The RP200930 data are fully consistent with this
period, but cannot be used to determine a more precise period because the
cumulative period uncertainty over the 6 months between observations greatly
exceeds the period.

The data from observation RP200930 are more suited to detailed time analysis.
In 1992 September the intensity was low, at a rate of 0.12~counts~s$^{-1}$
(Figure~3). One significant flare is superposed on the light curve. The source
is not constant. The character of the emission changed dramatically during the
second observation. The source was fairly constant during the first OBI.
Following an 8.3~Ksec gap in the observation, the source became highly variable
(Figure~4). The mean count rate quadrupled. Significant variations exist to
timescales shorter than 2~seconds; on this scale the instantaneous count rate
reaches 15~counts~s$^{-1}$.

We verified that this extreme variability was indeed a property of the X-ray
source, and not a detector artifact, by examining the light curves of other
X-ray sources in the field. None exhibit significant variability during this
interval. The background count rate in the annulus surrounding the source is
similarly constant. Because the source is well off-axis, the point spread
function overlaps several of the wires of the coarse window support mesh at all
times. The spacecraft wobble has no significant effect on the observed light
curve.

We searched for periodicities within the individual OBIs using FFT and
period-folding techniques. No significant periodicities were found on
timescales of a few seconds to 1000 seconds.

\subsection{The X-ray Spectrum}

We binned the background-subtracted PI (pulse height invariant) data from
RP200930 into the 34 SASS channels. After excluding channels 1-2 and 33-34, we
fit the source counts using XSPEC version 8.40. The fits are summarized in
Table~\ref{tbl-spec}, and shown in Figure~5. We found acceptable fits for a
blackbody spectrum with kT=50~eV or a powerlaw spectrum with $\alpha$=8. We
found no acceptable fit for a thermal plasma. The spectrum is highly absorbed,
which suppresses much of the uncertainty due to gain calibration in the lowest
energy channels. In the following discussion, we will refer only to the
blackbody spectral fits, because the soft X-ray emission from cataclysmic
variables is expected to, and indeed is observed to be, blackbody in character
(e.g., Lamb 1985).

We separately fit the data obtained when the the source was ``quiescent'' and
when the source was ``active''. There was no significant difference in either
the soft X-ray temperature or the absorption column. When the source was
active, we subdivided the data into a bright source (instantaneous count rates
$>$2~c~s$^{-1}$) and a faint source; again, there were no significant
differences in the spectral fits.

We did not attempt to fit the data from RP700422 because of the extreme
distance off-axis. A visual inspection of the PI pulse-height distribution
shows no significant differences from that of the data actually fit.

We see no evidence of a hard X-ray bremsstrahlung component in these data. We
fit the spectrum with two spectral components, a black body component plus a
hard bremsstrahlung component. The inclusion of the hard component led to a
modest reduction in $\chi^2$ (Table~\ref{tbl-spec}), but there was no
significant change in the blackbody temperature, and the bremsstralung
temperature and normalization are unconstrained. Over the energy range the PSPC
is sensitive to, we can place no meaningful limits on either the hard X-ray
flux or the ratio of ${L_{bb}}/{L_{brems}}$ in this object.

\section{The Optical Counterpart}

A finding chart is presented in Figure~6. The optical counterpart (marked) is
the westernmost and brighter of two stars near the X-ray position; it is not in
the HST Guide Star Catalog (Lasker \etal 1990). We determined the J2000 optical
position, 5$^h$15$^m$41.42$\pm$0.01$^s$ +1$^o$4'40.7$\pm$0.4", relative to 3
nearby stars in the guide star catalog. The star is about 18~arcsec from the
weighted mean X-ray position. There is a fainter star about 15~arcsec to the
east. The target was optically identified by its spectrum.

\subsection{The Optical Spectrum}

The optical spectra (Figures~7 and 8) were obtained using the KPNO 2.1m
telescope with the GOLDCAM camera and grating 47. The red spectrum was observed
on 1993 January 1. It covers the $\lambda\lambda$5500-7600\AA\ interval at
about 3\AA\ resolution. The blue spectrum was observed on 1994 January 29. It
covers the $\lambda\lambda$4000-5000\AA\ interval at about 1.5\AA\ resolution.
The data were reduced and flux-calibrated using the IRAF APEXTRACT and ONEDSPEC
packages. The spectra were extracted using variance-weighting within APALL.
Subsequent analysis has been done in IDL using the ICUR spectral analysis
package.

The red spectrum is the average of two 900~second exposures obtained under good
seeing with non-photometric conditions. The spectrum was placed on a relative
flux scale using the stars HD17520, HD86986, HD109995, and HD217086 as
standards. The continuum is quite blue. The strongest line is H$\alpha$, with
an equivalent width W$_\lambda$(H$\alpha$)=--30\AA. The line is asymmetric,
with emission extending some 40\AA\ towards the blue. The line can be fit as
two Gaussians, with $\sigma$=6 and 20\AA\ and the broad component centered
7\AA\ to the blue of the narrow component. The other prominent emission lines
are He~I~$\lambda$5876, 6678, and 7065\AA . The brighter two of these lines
have profiles similar to H$\alpha$. There is a prominent broad Na~D absorption
feature redward of He~I~$\lambda$5876. Line measurements are summarized in
Table~\ref{tbl-lines}.

The blue spectrum is underexposed. It is an 1800 second exposure obtained with
the target between 2 and 3 hours west of the meridian. The seeing was poor; the
night was far from photometric. The 3 individual exposures were averaged to
produce the final spectrum. The flux standard Feige 34 was used to place the
star on a relative flux scale. Three emission lines (H$\beta$, H$\gamma$, and
He~II $\lambda~4686$) are visible in the unsmoothed spectrum. Equivalent widths
of the lines are compiled in Table~\ref{tbl-lines}.

We did not obtain spectra of the faint star to the east.

\subsection{Optical Photometry}

Optical photometry was obtained on 3 separate photometric nights, 1993 December
5, 15, and 17 (Table~\ref{tbl-phot}). The observations on December 5 were made
from Kitt Peak using the 0.9 meter, the Bessell filters from the Harris filter
set and the T2KA CCD detector.  The later observations were made with the 31"
telescope on Anderson Mesa which is jointly operated by Lowell Observatory and
Northern Arizona University.  These observations were made with standard B
through I Bessell filters and a Photometrics liquid nitrogen cooled CCD. All
data were processed using IRAF for general debiasing and flat-fielding.  The
digital photometry was undertaken using IRAF with a mean extraction aperture
size of 5~arcsec. Photometry was calibrated by comparison with standards
measured by Landolt (1983). Magnitudes and colors are presented in
Table~\ref{tbl-phot}.

Taking the mean V magnitude to be 15.5, the $\frac{f_X}{f_V}$ ratio is about 6,
where $f_X$ is the integrated, dereddened flux from the 49~eV black body. When
faint, the system appeared redder. It is possible that the 5~December
observations caught an eclipse of the principle accretion pole, and that the
light is dominated by the secondary. The colors, however, are are not those of
a star, so some blue continuum must have been present.

Near IR Photometry was obtained on the 1993 October 30 using the Simultaneous
Quad Infrared Imaging Device (SQIID) on the 1.3 meter telescope at KPNO.  Three
180 second exposures were taken of the target.  Routines written by the authors
in IDL were used to subtract the dark current and sky backgrounds from the raw
frames as well as to preform the flat-fielding and coaddition to create final
frames. The near-infrared photometry was calibrated by comparison with
standards measured by Elias \etal (1982). The near-IR colors are consistent
with a hot blackbody (T$\gta$8000K), and not with emission from a cool
secondary. Note that the near-IR and optical photometry were obtained at
different epochs.

Further optical photometry has been obtained using the Stony Brook 14''
telescope with a Santa Barbara Instrument Group ST-6 CCD from 1994 March 18-25.
These data were obtained solely to look for evidence of eclipses, using 4 other
stars in the field as photometric standards. We observed a 7.6~x~5.7~arcmin
field. The average seeing at SUNY at Stony Brook is 3-5 arcseconds.  The images
consist of sixteen coadded frames with 15 second integration time each, giving
the final image an integration time of 4 minutes.  The images were flat-fielded
and dark-corrected using SBIG software.  IRAF was used for the photometric
measurements. No filters were used. The magnitude of the target is near the
practical limit for observing from Stony Brook, and because of the time of
year, the observations were made at high air mass (and in the direction of the
sky glow from New York City). We obtained 13 images over a 6 day span. Six
images were taken in a 110~minute interval on 1994 March 25; the other 7 images
were obtained between 1994 March 19 and 24. We see no convincing evidence for
variability at the $\pm$25\% level. We detected no eclipses. For a period of
$\sim$8~hours, we sampled essentially the same phases ($<$0.25 of the full
orbit) during the entire run.

\section{Discussion}

The optical and X-ray spectra, and the large $\frac{f_X}{f_{V}}$ ratio suggest
that this object is a magnetic cataclysmic variable (e.g., Mason 1985, Cropper
1990). The rapid X-ray variability requires a compact object. The relative
strengths of the He~II $\lambda$4686 and H$\beta$ emission lines (Liebert \&
Stockman 1985), and the strength of the soft-X-ray emission, are suggestive of
an AM~Her variable. The folded X-ray light curve, shown in Figure~2, is similar
to the X-ray light curve of VV~Pup (Osborne \etal 1984), though on a much
longer period. The light curve shows that the quiescent interval lasts about
half the period, followed by a smooth increase to a maximum about 7 times
brighter than the quiescent level. We assume this period is the orbital period
of the system. Based on the light curve, we consider it likely that this object
is a polar or an intermediate polar wherein the principle accreting pole is
occulted every stellar rotation.

No currently catalogued AM~Her system has a period longer than 4.6~hours
(Ritter \& Kolb 1993). If this is indeed an AM~Her system, with a
synchronously-rotating white dwarf, then the magnetic field strength must be
quite large to force circularization at an 8 hour orbital period. Using
Patterson's (1994) scaling law, B must be of order 7$\times$10$^8$G if the
accretion rate and the mass of the white dwarf are typical of magnetic
cataclysmic variables. Such a field strength, or even a considerably weaker
field, should produce easily-visible circular polarization. If the white
dwarf's rotation is not synchronous, then this object may represent the
progenitors of an AM Her-type system. This object may provide some insights
into the relation between the polars and the intermediate polars.

Recently, Garnavich \etal (1994) reported that the system is an eclipsing
magnetic cataclysmic variable, with a 7.98~hour orbital period. They estimated
a magnetic field strength of up to 5.5$\times$10$^{7}$~G from the strength of
the cyclotron emission humps. This is fully consistent with our
characterization based on the X-ray light curve. The magnetic field is weaker
than predicted from Patterson's scaling law, suggesting that either the white
dwarf is about half the mass of that in a typical magnetic cataclysmic
variable, that the system is not in synchronous rotation, or that the scaling
law breaks down for large separations.

The most unusual aspect of this object is the X-ray variability on short
timescales. Figure~9 shows a segment of the active interval during RP200930.
Note that the X-ray emission breaks up into a sequence of bursts with typical
duration of 10 seconds. There is no evidence for a steady component underlying
the bursts. The bursts are not periodic, but there is a mean separation time
between them of order 30~seconds. These appear comparable to the quasi-periodic
bursts seen in AM~Her by Tuohy \etal (1981), although in AM~Her the bursts are
superposed on a strong background. Short bursts have also been seen in BL~Hyi
(Beuermann \& Schwope 1989). The detected flux in the typical burst is
3$\times$10$^{-9}$~erg~cm$^{-2}$, yielding a burst energy \begin{center}
E$_b~\sim$~1.3$\times$10$^{34}$~D$^2_{200}$~erg \end{center} where D$_{200}$ is
the distance to the source in units of 200~pc. The mass of the accreting blob,
if all its kinetic energy is released in the soft X-rays, is
\begin{center}M$_{blob}$~=~1.0$\times$10$^{17}$D$^2_{200}$R$_9$/M$_1$~gm,
\end{center} where R$_9$ and M$_1$ are the radius and mass of the white dwarf
in units of 10$^9$~cm and one solar mass, respectively.

The soft blackbody component is generally thought to be attributable to thermal
reprocessing of harder bremsstrahlung and cyclotron X-ray emission. The
discrete bursts are suggestive of blobby accretion models developed by Kuijpers
\& Pringle (1982) and Frank \etal (1988). The blobs possess sufficient ram
pressure to plunge deep into the photosphere before shocking. The hard X-ray
bremsstrahlung radiation suffers multiple scatterings and emerges as the soft
blackbody component seen in many AM~Her systems (e.g., Lamb \& Masters 1979).

We can estimate the mean mass accretion rate, $\dot{\rm m}$, and the fractional
area, f, undergoing accretion by equating the blackbody luminosity
\begin{center}L$_{\rm BB}$~=~4$\pi$R$^2$f$\sigma$T$^4$,\end{center}
the observed X-ray luminosity (dereddened and extrapolated over the entire
wavelength range of the blackbody emission, using the spectral fit)
\begin{center}L$_X$~=~4$\pi$d$^2$f$_x$,\end{center}
and the accretion luminosity
\begin{center}L$_{\rm acc}$~=~GM$\dot{\rm m}$/R,\end{center}
where we assume that all the accretion luminosity is radiated into the
blackbody component.
Plugging in the observed temperature and observed X-ray flux,
L$_{\rm BB}$~=~7.5$\times$10$^{37}$R$_9^2$f,
L$_{X}$~=~7.5$\times$10$^{31}\frac{f_x}{<f_x>}$D$_{200}^2$, and
L$_{\rm acc}$~=~1.3$\times$10$^{17}$M$_1\dot{\rm m}$/R$_9$.
Rearranging these, we find that
\begin{center}
  f~=~1.0$\times$10$^{-6}\frac{f_x}{<f_x>}(\frac{D_{200}}{R_9})^2$,
  and\\
  $\dot{\rm
m}$~=~5.8$\times$10$^{14}\frac{f_x}{<f_x>}$D$^2_{200}\frac{R_9}{M_1}$~gm~s$^{-1}$,
\end{center}
where $f_x$ and $<f_x>$ are the instantaneous and mean X-ray fluxes.
For a typical luminosity of L$_{33}$~=~10$^{33}$erg~s$^{-1}$,
\begin{center}
   $\frac{f_x}{<f_x>}$D$^2_{200}$=13.3L$_{33}$,\\
   f~=~1.3$\times$10$^{-5}$L$_{33}$R$_9^{-2}$, and \\
   $\dot{\rm m}$~=~7.7$\times$10$^{15}$L$_{33}$R$_9$/M$_1$
\end{center}
The instantaneous luminosity varies by nearly two orders of magnitude, but we
see no evidence for significant changes in the temperature. Therefore,
increases in luminosity must be due to increases in $\dot{\rm m}$ and f. Both
scale linearly with the instantaneous flux. This suggests that all the blobs
penetrate to sufficiently large optical depths to thermalize completely. The
more massive blobs may penetrate deeper and illuminate a larger fraction of the
surface area.

The wide binary orbit can accomodate a relatively luminous secondary. The Na~D
absorption (Figure~7), coupled with the lack of strong TiO absorption bands,
suggests a secondary with a G to late-K spectral type. The Na~D absorption
could be partly interstellar, but are the strongest metallic features in the
red in G-K stars; our spectrum is too noisy to detect the $\lambda$6495\AA\
blend. The colors observed on 1993 December 5 are fairly red, but are not
consistent with any cool star. If the secondary has the colors of a normal
dwarf, and the light is the sum of the secondary and a blue continuum, then the
secondary cannot be earlier than about spectral type K2. Garnavich \etal (1994)
obtained an M0 spectral type for the secondary from a spectrum during eclipse.

This system provides strong evidence for accretion through discrete blobs. In
this system the blobby accretion may not be just sporadic, as has been seen in
other polars, but may be the normal state. Accretion is not expected to produce
a smooth hydrodynamic flow, and larger separation between components may give
more time for the instabilities to grow and to produce better-defined blobs
than in the shorter-period systems. If the fraction of the accreting matter in
discrete blobs is large, as suggested by the X-ray light curve, then the hard
X-ray flux from this object should be minimal.

In conclusion, RXJ051541+0104.6 is a magnetic cataclysmic variable with a
period over twice as long as the longest known AM~Her system. Further studies
may reveal much about both the accretion processes in close binaries and the
evolution of cataclysmic variable systems.

\acknowledgments
We thank M. Watson and J. Patterson for beneficial discussions. An anonymous
referee offered a number of useful suggestions. We thank P. Szkody for sending
us her results in advance of publication. The photometric observations from
Stony Brook were obtained by undergraduate observers  Matthew Adcock, James
Petreshock, Scott Schell, and Daniel O'Sullivan. The observing time at the
Lowell Observatory was granted through the generosity of K. Eastwood. We
acknowledge the superlative efforts of the ROSAT personnel in Germany and at
the NASA Goddard Space Flight Center in making the ROSAT observatory a
scientific success. This investigation has been supported by NASA grant
NAG5-1663.

\clearpage
\begin{table*}
\begin{center}
\caption{X-ray Observation Log }\label{tbl-x}
\begin{tabular}{lrlrrrr}
Observation &\multicolumn{1}{c}{Pointing}&date& start & stop &
     N\tablenotemark{a} &seconds\\
            & \multicolumn{1}{c}{(J2000)}  &&\multicolumn{2}{c}{UT} & \\
\tableline
RP700422    & 5 16 11.0 +0 09 0.1 & 1991 Sept 18 & 16:44
            &16:33\tablenotemark{b}&35&24493\\
            & \\
RP200930    & 5 14 24.0 +1 42 0.0 & 1992 Sept 11 & 19:39 & 20:05 & 1& 1488\\
            &                     & 1992 Sept 19 &  5:39 &  5:38 & 2& 2165\\
            &                     & 1993 Feb  27 & 12:22 & 16:25 & 3& 6126\\
\tableline
\end{tabular}
\end{center}
\tablenotetext{a}{Number of OBIs containing good data.}
\tablenotetext{b}{Observation ended on 1991 September 19.}
\end{table*}

\clearpage
\begin{table*}
\begin{center}
\caption{Summary of Spectral Fits }\label{tbl-spec}
\begin{tabular}{lrrrrr}
model       & n$_H$                & P1\tablenotemark{a} & kT$_{brems}$
           & $\chi^2$ & DoF\tablenotemark{b} \\
            &cm$^{-2}$ &  \\
\tableline
blackbody   & 5.4$\pm$1.3$\times$10$^{20}$ &  49$\pm$5  & --- & 30.9 & 27 \\
power law   & 1.0$\pm$0.2$\times$10$^{21}$ & 8.2$\pm$1.1& --- & 25.4 & 27 \\
            & \\
bb+brems    & 6.0$\pm$1.6$\times$10$^{20}$ & 46$\pm$6 &13$\pm$273 & 25.8 & 25\\
\end{tabular}
\end{center}
\tablenotetext{a}{Spectral fit parameter: kT (ev) for the blackbody fit, or
                  $\alpha$ for the power law fit.}
\tablenotetext{b}{Degrees of freedom for the spectral fit.}
\end{table*}

\clearpage
\begin{table*}
\begin{center}
\caption{Summary of Line Measurements }\label{tbl-lines}
\begin{tabular}{lrrr}
line & $\lambda$ & date & W$_\lambda$(\AA)\\
\tableline
H I  & 4340 & 29 Jan 1994 & -25:\\
H I  & 4860 & 29 Jan 1994 & -20:\\
H I  & 6563 &  1 Jan 1993 & -30 \\
He I & 5876 &  1 Jan 1993 & -5. \\
He I & 6678 &  1 Jan 1993 & -5. \\
He I & 7065 &  1 Jan 1993 & -3.5 \\
He II & 4686 & 29 Jan 1994 & -22:\\
Na I  & 5890 & 1 Jan 1993 & 3.3 \\
\end{tabular}
\end{center}
\end{table*}

\clearpage
\begin{table*}
\begin{center}
\caption{Summary of Photometric Observations}\label{tbl-phot}
\begin{tabular}{lrrrrr}
Date (UT)          & \multicolumn{1}{c}{V} & \multicolumn{1}{c}{U-B} &
  \multicolumn{1}{c}{B-V} & \multicolumn{1}{c}{V-R} & \multicolumn{1}{c}{R-I}\\
\tableline
1993 December 5.40 & 16.05$\pm$.03& -0.42$\pm$.06& 0.61$\pm$.06& 0.49$\pm$.04
                                            & 0.48$\pm$.05\\
1993 December 15.21 & 15.58$\pm$.02& --- & 0.21$\pm$.04& 0.52$\pm$.02
                                            & 0.24$\pm$.04\\
1993 December 17.21 & 15.47$\pm$.02& --- & 0.31$\pm$.05& 0.30$\pm$.02
                                            & 0.42$\pm$.02\\
\tableline
Date (UT)          & \multicolumn{1}{c}{K} & \multicolumn{1}{c}{J-K} &
  \multicolumn{1}{c}{H-K} \\
\tableline
1993 October 30.32 & 13.94$\pm$.06& 0.01$\pm$0.06 & 0.00$\pm$0.02 \\

\end{tabular}
\end{center}
\end{table*}

\newpage
\begin{center}{Figure Captions}\end{center}

\skp
Figure 1. The soft X-ray curve of RJ051540+0104.6 during the
1991 September observation (RP700422). The data are binned in 800 second bins.
Background has been subtracted.
The mean count rate is indicated by the dashed line.
The source was near the edge of the field, and wobbled out of the field of view
for 100 of every 400 seconds (see text); these count rates have not been
corrected for this effect. The overall light curve suggests about an 8~hour
periodicity.

\skp
Figure 2. The X-ray flux from RP700422 (Figure~1) folded on an 8.05~hour
period.
Two periods are plotted for clarity; zero phase is arbitrary.

\skp
Figure 3. The ``quiescent'' light curve during the first 5 OBIs of observation
RP200930 from 1992 September and 1993 February
27. The background-subtracted, vignetting-corrected
count rate is plotted in 100 second time bins. The short dashed horizontal
line represents the mean count rate.
The two dashed vertical lines represent data gaps of approximately 8 days and
5.5 months, respectively. The source is variable.

\skp
Figure 4. The erratic light curve seen during the last OBI of observation
RP200930, in 30~second bins. The source is highly variable,
with significant stretches of zero intensity.

\skp
Figure 5. The X-ray spectrum from RP200930 with the best fit blackbody
          spectrum (kT=49~eV) overplotted.

\skp
Figure 6. Finding chart.
This image is a 20~second V-band observation taken with the KPNO 0.9m
on 1993 December 5.  The pixel size is $\approx$~0.7~arcsec. The figure
is  $\approx$~4.7~arcmin on a side. North is up and east is to the left.
The cataclysmic variable is marked.

\skp
Figure 7. The blue spectrum, with a resolution of about 1.5\AA, after
smoothing with a 2\AA\ FWHM Gaussian. The
prominent emission lines are H$\gamma$, He~II~$\lambda$4686, and H$\beta$.
The narrow spikes atop H$\gamma$ and at 5000\AA\ may be due to cosmic rays.

\skp
Figure 8.
The red spectrum, with a resolution of about 3\AA. Note the blue
continuum. He~I and H$\alpha$ emission is prominent. Na~D is in absorption,
but no other features attributable to a cool secondary are visible. If Na~D
is from a cool secondary, the spectral type cannot be later than mid-K because
of the absence of TiO absorption bands.

\skp
Figure 9. The X-ray bursts seen in part of the observation shown
in Figure~4. The data are binned into 2~second time bins. Error bars are
omitted for clarity (after the vignetting correction, single photons in a
2~second bin contribute
0.7~counts~s$^{-1}$). Negative excursions are subtracted background photons.
The mean count rate (the dashed horizontal line) corresponds to just under one
photon per time bin.
The bursts appear symmetric,
with durations $<$10~sec and peak fluxes over 2 orders of magnitude above the
``quiescent'' level. The bursts are not superposed on a steady source.


\clearpage

\begin{references}
Aschenbach, B. 1988, Appl. Opt., 27(8), 1404.

Beuermann, K. \& Schwope, A. D. 1989, A\&A, 223, 179.

Beuermann, K. \& Schwope, A. D. 1994, in Interacting Binary Stars (ASP
     Conference Series, Vol 56), ed.
     A. W. Shafter, (San Francisco: Astronomical Soc. Pacific), 119.

C\'{o}rdova, F. 1993, in X-Ray Binaries, eds.
      W. H. G. Lewin, J. van Paradijs, and E. P. J. van den Heuvel
     (Cambridge:Cambridge University Press)

Cropper, M. 1990, Space Sci. Rev., 54, 195.

Elias, J. H., Frogel, J. A., Matthews, K., \& Neugebauer, G. 1982, AJ, 87,
     1029.

Frank, J., King, A. R., \& Lasota, J.-P. 1988, A\&A, 193, 113.

Garnavich, P. M., Szkody, P., Robb, R. M., Zurek, D. R., \& Hoard, D. 1994,
     submitted to ApJL.

Kuijpers, J. \& Pringle, J. E. 1982, A\&A, 114, L4.

Landolt, A. U. 1983, AJ, 88, 439.

Lamb, D. Q., \& Masters, A. R. 1979, ApJ, 234, L117.

Lamb, D. Q., 1985, in Cataclysmic Variables and Low Mass X-Ray Binaries,
     eds. D. Q. Lamb and J. Patterson (Reidel: Dordrecht)  p. 179.

Lasker, B. M., Sturch, C. R., McLean, B. J., Russell, J. L, Jenker, H.,
        \& Shara, M. M. 1990, AJ, 99, 2019.

Liebert, J. \& Stockman, H. S. 1985, in Cataclysmic Variables and Low
     Mass X-Ray Binaries,
     eds. D. Q. Lamb \& J. Patterson (Reidel: Dordrecht)  p. 151.

Mason, K. O. 1985, Space Sci. Rev., 40, 99.

Osborne, J. \etal 1984, in X-ray Astronomy 84, ed. Y. Tanaka.

Patterson, J. 1994, PASP, 106, 209.

Pfeffermann, E. \etal 1987, in Soft X-Ray Optics and Technology, ed. E.-E. Koch
     \& G. Schmahl (Proc. SPIE, 733), 519.

Ritter, H. \& Kolb, U. 1993, in X-Ray Binaries, eds. W. H. G. Lewin, J. van
     Paradijs, \& E. P. J. van den Heuvel, (Cambridge:Cambridge University
Press).

Tr\"umper, J. 1983, Adv Space Res., 2(4), 241.

Tuohy, I. R., Mason, K. O., Garmire, G. P., \& Lamb, F. K. 1981, ApJ, 245, 183.

Walter, F. M. 1994, in Proceedings of the ROSAT Science Workshop, Ed.
E. Schlegel (in press).

\end{references}
\end{document}